# 2-aminooxazole in astrophysical environments: IR spectra and destruction cross sections for energetic processing.

Belén Maté[1*], Ricardo Carrasco-Herrera[1], Vicente Timón[1], Isabel Tanarro[1], Victor J. Herrero[1],

Héctor Carrascosa[2], Guillermo M. Muñoz Caro[2], Cristóbal González-Díaz[2], Izaskun Jiménez-Serra[2].

1. Instituto de Estructura de la Materia, IEM-CSIC, Serrano 121-123, 28006 Madrid, Spain.
2. Centro de Astrobiología, INTA-CSIC, Carretera de Ajalvir, km 4, Torrejón de Ardoz, 28850 Madrid, Spain.

*belen.mate@csic.es

**Abstract**

2-aminooxazole (2AO), a N-heterocyclic molecule, has been proposed as an intermediate in prebiotic syntheses. It has been demonstrated that it can be synthesized from small molecules such as cyanamide and glycoaldehyde, which are present in interstellar space. The aim of this work is to provide infrared spectra, in the solid phase for conditions typical of astrophysical environments and to estimate its stability toward UV photons and cosmic rays. Infrared (4000-600 $cm^{-1}$) absorption spectra at 20 K, 180 K, and 300 K, IR band strengths, and room temperature UV (120-250 nm) absorption spectra are given for the first time for this species. Destruction cross-sections of ≈ 9.5 $10^{-18}$ $cm^2$ and ≈ 2 $10^{-16}$ $cm^2$ were found in the irradiation at 20 K of pure 2AO and 2AO:H2O ices with UV (6.3-10.9 eV) photons or 5 keV electrons, respectively. These data were used for the estimate of half-life times for the molecule in different environments. It is estimated that 2AO could survive UV radiation and cosmic rays in the ice mantles of dense clouds beyond cloud collapse. In contrast, it would be very unstable at the surface of cold Solar System bodies like Kuiper belt objects, but the molecule could still survive within dust grain agglomerates or cometesimals.

## 1. Introduction

It is currently unknown what processes led to the emergence of life on Earth. One of the proposed scenarios is a primordial ribonucleic acid (RNA)-world, in which RNA molecules may have proliferated before the appearance of deoxyribonucleic acid (DNA) and proteins. RNA molecules indeed present the basic functionalities critical for life: they store genetic information, they self-replicate, and they catalyse chemical reactions essential for life. The theory of a primordial RNA-world was initially received with scepticism because early pre-biotic experiments revealed that it was far from trivial to form RNA from its basic compounds, a nitrogenous base, a ribose sugar and a phosphate. The bottleneck in this process was the addition of nucleobases to ribose, which not only is an inefficient reaction for purine nucleobases (Adenine and Guanine; Fuller et al. 1972) but it is also an inoperative mechanism for pyrimidine nucleobases (Thymine, Cytosine, Uracil; Orgel 2004).

In the past decade, however, a series of works have demonstrated that this bottleneck can be bypassed thanks to the formation of an intermediate N-heterocyclic compound called 2-aminooxazole ($C_3H_4N_2O$; Powner et al. 2009; Patel et al. 2015). Under the typical geo-chemical conditions of an early-Earth, this compound can be synthesized from small molecules such as cyanamide ($NH_2CN$) and glycolaldehyde ($HOCH_2CHO$), which are present in interstellar space (Turner et al. 1975; Hollis et al. 2000). If brought by meteor/cometary impacts to the surface of an early Earth, these molecules could have reacted in aqueous solution producing 2-aminooxazole (Powner et al. 2009).

But the question that follows is: could a N-heterocyclic molecule such as 2-aminooxazole (2AO) form already in interstellar space? Laboratory experiments have shown that this type of compounds can appear in interstellar ice analogues after being irradiated with UV-photons and after being thermally processed (see Meierhenrich et al. 2005, Oba et al. 2019). N containing heterocycles have also been detected in carbonaceous chondrites, which contain a large fraction of material from the pristine solar nebula, but they are usually assumed to form through chemical alteration at T > 100 K (Naraoka et al. 2017, Vinogradoff et al. 2020)

Deep searches of 2-aminooxazole have been carried out recently toward two of the most chemically-rich sources in our Galaxy, the low-mass warm core (hot corino) IRAS16293-2422 (e.g. Jørgensen et al. 2016) and the quiescent Giant Molecular Cloud G+0.693-0.027 located in the Galactic Center (Requena-Torres et al. 2008; Zeng et al. 2018). These searches did not yield any detection (Jiménez-Serra et al. 2020). From the high-sensitivity spectra toward IRAS16293-2422 and G+0.693-0.027, upper limits to the abundance of 2AO of ~ $1\times10^{-11}$ and ~$8\times10^{-11}$, respectively, were inferred. In contrast, urea, another key prebiotic precursor (Menor-Salván et al. 2009), was found toward the quiescent cloud G+0.693-0.027 with a derived abundance of $4.7\times10^{-11}$ (Jiménez-Serra et al. 2020). Although the upper limits to the abundance of 2AO are still high, this molecule may be more sensitive to highly energetic phenomena (such as UV-photon radiation or cosmic rays) than urea, explaining the non-detection of this N-heterocyclic species within the same cloud. In fact, G+0.693-0.027 is located in a hostile environment within the Central Molecular Zone (CMZ) of the Galactic Center, dominated by the interaction of large-scale shocks, intense UV-radiation fields and/or enhanced rates of cosmic rays (Goto et al. 2014; Zeng et al. 2018). Although the formation of complex organic molecules (COMs), like those observed in hot cores, is easier at temperatures where reactive species can diffuse within the ices (T > 20-30 K), COMs have also been observed in cold dense cores, at temperatures as low as 10 K (Jimenez-Serra et al. 2016 and references therein). Production of reactive species through UV or CR irradiation and non-diffusive reaction mechanisms (see discussion in Jin & Garrod 2020) can possibly account for COM formation within the ices of cold cores.

The photo-chemistry of 2-aminooxazole has been investigated, both theoretically (Szabla et al. 2013, 2015) and experimentally in aqueous solution at room temperature (Todd et al. 2019). However, nothing is known experimentally about its possible photo-products and the efficiency of this process under conditions relevant for astronomical environments. Complex organic molecules in the interstellar medium are generally assumed to form through ice chemistry. In this work, we study experimentally the photo-chemistry of 2-aminooxazole ices under the effects of both UV-photons and cosmic-ray radiation simulating interstellar conditions in the laboratory. The products of the photo-destruction of 2AO have been investigated, and its destruction cross sections (with both UV photons and 5 keV electrons) have been estimated. To the best of our knowledge, this work represents the first study of the photo-chemistry of a N-heterocyclic molecule with a H substituted by a functional ($NH_2$) group performed under conditions similar to those found in interstellar space. Our experiments have also allowed us to record for the first time the IR spectra of 2-aminooxazole at low temperatures for the wavelength range between 2.8 and 20 μm and therefore, these spectra may help in the identification of this prebiotic molecule in astrophysical ices with the next generation facilities such as the James Webb Space Telescope (JWST).

## 2. Experimental section.

The high-vacuum (HV) experimental setup employed for the study of 2-aminooxazole has been described previously (Maté et al. 2014, Maté et al., 2017), although some modifications have been implemented that need to be detailed. A new cryostat, ARS DE-204AB, has been installed on the top flange of the HV chamber. It is provided with a thermofoil heater and a silicon diode placed at the end of the cold head that, by means of a Lakeshore temperature controller, allows a 0.5 K accuracy temperature control of the end tip. A new copper sample holder, designed to work in a transmission configuration, has been installed in its cold head. It holds a 25 mm x 1mm Si wafer that leaves a circle of 12 mm diameter of its surface exposed. With the new cryostat and sample holder the end tip reaches 10 K in 45 min. The rest of the setup remains unchanged. The vacuum system provides a background pressure in the $10^{-8}$ mbar range at room temperature and close to $1 \times 10^{-8}$ mbar with the cryostat on. This background corresponds mostly to water vapour desorbed from the wall surfaces. In the ice deposition experiments, background water deposits on cold surfaces at an approximate rate of 0.1 nm $min^{-1}$. The chamber is coupled through KBr windows to a Vertex70 FTIR spectrometer. A rotatable flange allows the orientation of the Si surface to face the sublimation oven, the IR beam of the spectrometer, or the energetic processing equipment (UV lamp or electron gun). In the present work, the spectra have been recorded in normal transmission configuration, with a 2 $cm^{-1}$ resolution and averaging 200 scans.

At room temperature, 2AO is a light yellow powder with a molecular weight of 84.08 and a density of 1.24 g/$cm^{-3}$ (www.chemicalbook.com) commercially available (97%, Sigma-Aldrich) that has to be stored at 2-8 ºC, and does not need special safety care. Its melting point is 90-95 ºC. To sublimate this species, the high-vacuum-sublimation oven described in a previous work (Maté et al., 2014), has been employed. However, in the first tests it was found that 2AO reacts with the copper wall of the oven crucible. It is transformed into an orange solid with a lower vapour pressure, and the IR spectrum of the material deposited on the substrate does not correspond to 2AO. To overcome this problem, the internal wall of the sample compartment was covered with a stainless steel foil, which proved to be inert. Then, the oven was placed in the HV chamber, with its 3 mm diameter exit hole at a distance of 30 mm from the cold Si surface. The oven was heated to 50ºC, to increase the vapour pressure of 2AO. When the Si substrate had reached the desired temperature, it was rotated to face the oven, and the oven shutter was opened. Most of the 2AO vapour leaving the crucible hit the cold Si surface and froze generating a layer. Deposition times of several minutes were typically used. The deposition of thin layers (< 200 nm) was difficult with this set-up.

Ice mixtures of 2AO:$H_2O$ were generated by introducing simultaneously both vapours into the vacuum chamber. The water vapour line, provided with a leak valve, is placed in a port that ensures backfilling of the chamber. Therefore, during the simultaneous deposit, a mixed ice layer grows on one side of the Si substrate (the one facing the oven) and a pure water ice layer accretes on its back side. In these experiments, a spectrum of pure amorphous solid water (ASW) was subtracted from the measured spectra, to analyse only the contribution of the side containing the mixture.

Two sets of experiments were performed. On the one hand, the infrared spectrum of 2AO was studied under astrophysical conditions. For this purpose, 2AO ices, pure and mixed with water, were grown at 20 K and warmed to monitor phase changes in the sample. On the other hand, 2AO ices, pure and mixed with water, were generated at 20 K and processed at this temperature with UV photons or 5 keV electrons to obtain information about the stability of this species in different astrophysical environments.

For UV processing, a Hamamatsu L10706 $D_2$ lamp, whose emission is mostly concentrated at wavelengths between 120 and 180 nm (10.3 -6.9 eV) has been employed. It provides a photon flux of 7.5 x $10^{13}$ photons $cm^{-2}$ $s^{-1}$ that is spatially homogeneous over the whole 12 mm diameter exposed surface of the Si substrate (located at 30 mm from the lamp). This flow was estimated from the irradiance of the lamp provided by the manufacturer, integrating in the 120-180 nm wavelength range, as described in Maté et al. (2018). The source of high energy electrons is an electron gun built in our laboratory (Maté et al., 2014). The gun provides a homogeneous and stable electron flux that, like the UV lamp, covers the 12 mm diameter exposed area of the Si substrate. For this work, a flux of 4 x $10^{12}$ electrons $cm^{-2}$ $s^{-1}$ has been employed. The gun was calibrated by measuring the current reaching a conducting 12 mm diameter disk, placed at the position of the Si wafer.

Apart from the experiments detailed above, other experiments were performed to determine IR absorption band strengths and the UV absorption spectrum of solid 2-aminooxazole at room temperature.

*IR absolute band strengths*

As far as we know, no literature values have been reported for 2AO IR band strengths. In this work, they have been estimated from the IR spectra, between 4000 $cm^{-1}$ and 500 $cm^{-1}$, of KBr pellets containing a known number of 2AO molecules. Initially 2.5 ± 0.3 mg of 2AO were weighted and mixed with KBr to form four KBr pellets. The spectra of these four pellets were added. A column density of 1.35 ± 0.2 x $10^{19}$ molec /$cm^2$ was estimated for a 13 mm diameter disk pellet that contains 2.5 mg of 2AO. From the law of Lambert-Beer, the absorption strength of an infrared band, A', is given by: $A' = 2.303\ Int/N$ , where *Int* is the integrated area of the band in the absorbance spectrum and *N* the column density.

The thickness of the 2AO ice layers generated in this work was estimated from its IR transmission spectrum with the band strengths, measured as described in the previous paragraph, and with the density given above (1.24 g/cm$^3$). The band at 1423 cm$^{-1}$ was selected for this estimation. It was assumed that the room temperature band strengths of the polycrystalline solid are approximately valid for the amorphous ice. We are aware that this is a crude approximation but have made it to have an estimate of the band strengths, given the absence of experimental values in the literature. The estimated thickness of the ice layers grown varies between ≈ 60 and 550 nm.

UV absorbance spectrum

The vacuum-ultraviolet (VUV) absorption of 2AO was measured at the interstellar Astrochemistry Chamber (ISAC), described in Muñoz Caro et al. (2010). Basically, ISAC consists of an ultra-high-vacuum (UHV) chamber, with pressure typically in the range P = 3–4 x 10$^{-11}$ mbar, where an ice layer is deposited at 8 K. The 2AO sample was characterized by transmittance FTIR spectroscopy and VUV-spectroscopy. The source used for VUV-spectroscopy is an F-type microwave-discharge hydrogen flow lamp (MDHL), from Opthos Instruments, with a VUV-flux of 2.5 x 10$^{14}$ cm$^{-2}$ s$^{-1}$ at the normal sample position measured with a calibrated Ni-mesh.

The VUV-spectrum was measured using a McPherson 0.2 m focal length VUV monochromator (model 234/302) with a photomultiplier tube (PMT) detector equipped with a sodium salicylate window, with a resolution of 0.2 nm (see Cruz-Díaz et al. 2014a for details).

The main emission bands are Ly-α at 121.6 nm (10.20 eV) and the molecular hydrogen bands around 157.8 nm (7.85 eV) and 160.8 nm (7.71 eV) for the operating hydrogen pressure of 0.4 mbar. The interface between the MDHL and the vacuum chamber is a MgF$_2$ window. The monochromator is located at the rear end of the ISAC chamber, separated by another MgF$_2$ window. Due to its high volatility under vacuum, the 2AO sample was not placed inside the ISAC chamber, instead it was located close to the entrance slit of the VUV monochromator, pumped up by a turbo molecular pump. The background VUV-spectrum corresponding to the MDHL lamp emission was subtracted from the one with the 2AO sample. The sample was sandwiched in two MgF$_2$ windows to reduce desorption. Measurement of the IR absorption of 2AO samples did not allow a proper estimation of the column density due to the variations of sample thickness, which must have also affected the quality of the VUV-measurements. For this reason, no estimation of the VUV-absorption cross section is provided.

*Theoretical method. Crystalline structure and infrared spectrum.*

A tentative solid state theoretical structure of 2AO was built with first-principles solid-state methodology. An infrared spectrum for the solid was derived using this theoretical structure. The calculated spectrum was compared to the measurements and used for vibrational mode assignment. Periodic DFT calculations were performed with the use of Cambridge Serial Total Energy Program (CASTEP) (Clark et al. 2005) on a tentative 2AO molecular ice crystal model since, as far as we know, no crystallographic data have been reported for this molecule. To construct the periodic-simulation unit cell, the Amorphous Cell module in the Materials Studio suite was employed. In this way, it was possible to build a model of a 2AO homogenous 3D ice crystal structure composed by six molecules. An image of the distribution of molecules in the unit cell is given in Figure 1.

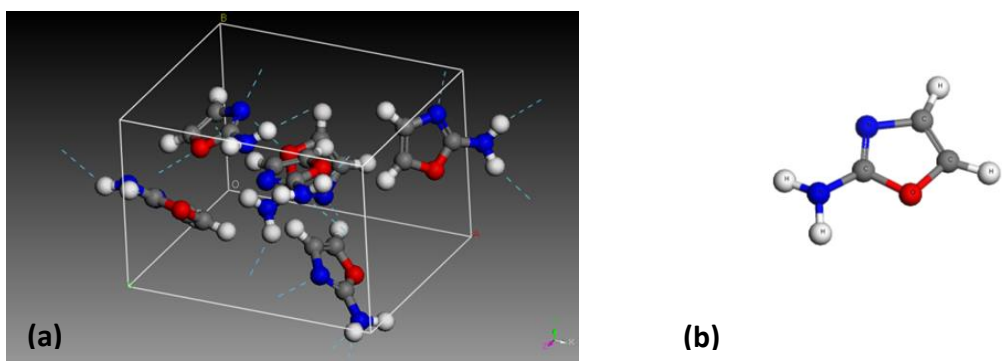

(a) (b)

Figure 1. (a) Unit cell of the computed structure of the 2-aminooxazoleO molecular crystal (blue dashed lines show the hydrogen bond network). (b) Monomer of 2-aminooxazole

All the computations were performed with the Perdew-Burke-Ernzerhof (PBE) energy-density functional (Perdew et al., 1996) complemented with Grimme's empirical dispersion correction (Grimme, 2006) (DFT-D2 approach). The specific pseudopotentials employed were standard norm-conserving pseudopotentials (Payne et al., 1992) from the CASTEP package. The amorphous created cubic unit-cell parameters and the associated atomic positions were fully optimized by means of the Broyden–Fletcher–Goldfarb–Shanno (BFGS) technique (Pfrommer et al., 1997). A plane wave kinetic energy cut-off parameter of ε = 830 eV and a dense $k$-mesh of 1× 2 × 2 (2 $k$-points) were employed. The lattice constants obtained were: a= 10.16 Å , b= 8.33 Å, c= 8.14 Å,  α= 85.58 $^0$, β= 100.24 $^0$, γ= 8.14 $^0$. With this well converged crystal structure, density functional perturbation theory (DFPT) (Baroni et al., 2001) (Refson et al. 2006) provides an analytical way of computing the second derivative of the total energy with respect to a given perturbation. The harmonic vibrational frequencies are obtained from the matrix of Cartesian second derivatives, also known as the Hessian matrix, of a molecular or periodic system (Wilson, 1955). Intensities of a given mode can be evaluated as a square of all transition moments of this mode and expressed in terms of the Born effective charge matrix and eigenvectors of the mass-weighted Hessian. The periodic DFT approach was originally developed to deal with inorganic systems but can also predict the vibrational spectra of molecular crystals with very good accuracy (Derringer et al. 2017). Lattice expansion can strongly perturb collective vibrations below 300/400 cm$^{-1}$, but has a much smaller influence on the mid-IR frequencies calculated in our spectra. The method has been successfully used in previous work by our group for the calculation of IR spectra of a molecular crystal (Maté et al. 2017)

## 3. Results and Discussion

*3.1. Room temperature and simulated 2-aminooxazole infrared spectra*.

The spectrum of 2-aminooxazole powder embedded in a KBr pellet is presented in Figure 2 (top trace). It corresponds to the average of the spectra of the four pellets described in section 2. The IR band strengths, A', obtained from this spectrum are given in Table 1. The estimated uncertainty is about 25 %, due mainly to the inaccuracy in the molecular number density.

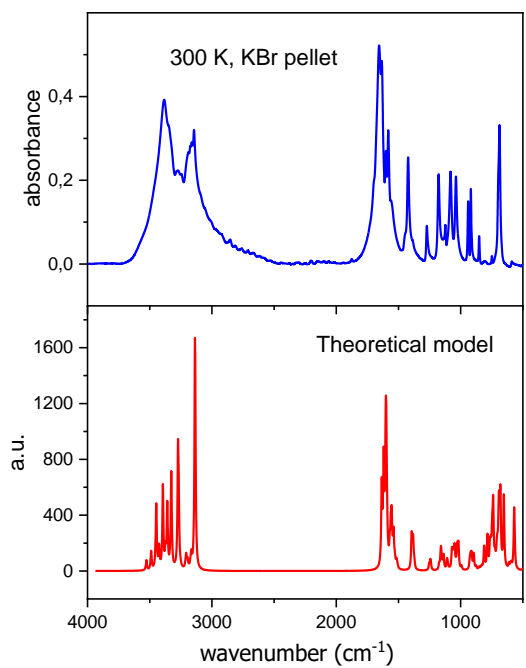

Figure 2. IR spectra of 2AO KBr pellet (top panel) and of the simulated crystal of this species (bottom panel). The simulated absorption features have been broadened with Gaussians of 10 cm$^{-1}$ FWHM. It corresponds to a 2AO column density of $1.35 \times 10^{19}$ molec cm$^{-2}$.

Table 1. IR absorption wavenumbers (cm$^{-1}$) and band strengths for the experimental and simulated absorptions of 2-aminooxazole. Assignment of the bands to different 2AO molecular groups, guided by the simulations, are given in the last column. Symbols ν, β, δ, ρ and ω stand for stretch, deformation, bending, rocking and wagging vibrations respectively. The subindex a denotes antisymmetric mode. Uncertainties in the A' are of the order of 25%, due mostly to the inaccuracy in the estimation of the column density.

| Peak (cm$^{-1}$) | Wavelength range (cm$^{-1}$) KBr pellet | A' (cm molec$^{-1}$) KBr pellet | Wavenumber range (cm$^{-1}$) theoretical | A' (cm molec$^{-1}$) theoretical | band assignment |
|---|---|---|---|---|---|
| 3133 | 3660-2500 | 9.7 x 10$^{-17}$ | 3600-3000 | 2.5 x 10$^{-16}$ | ν NH$_2$ |
| 1655,1592 | 1930-1473 | 4.0 x 10$^{-17}$ | 1770-1450 | 1.7 x 10$^{-16}$ | δ NH$_2$ |
| 1423 | 1473-1350 | 6.6 x 10$^{-18}$ | 1450-1303 | 2.3 x 10$^{-17}$ | ν$_a$ OCN |

| 1273 | 1293-1220 | 1.6 x 10^-18 | 1300-1200 | 6.5 x 10^-18 | β δ ring |
| --- | --- | --- | --- | --- | --- |
| 1173 | 1213-1115 | 5.2 x 10^-18 | 1200-1090 | 2.0 x 10^-17 | δ NCH |
| 1088 | 1115-1056 | 4.5 x 10^-18 | 1090-1040 | 1.8 x 10^-17 | ρ NH$_2$ |
| 1039 | 1056-1007 | 3.5 x 10^-18 | 1040-965 | 1.6 x 10^-17 | ρ CH |
| 944, 916 | 960-892 | 2.7 x 10^-18 | 965-863 | 1.7 x 10^-17 | ω HCCH + ω NH$_2$ |
| 850 | 870-837 | 4.9 x 10^-19 | 863-623 | 1.4 x 10^-16 | ρ HCCH |
| 701 | 740-630 | 5.3 x 10^-18 | | | |

The experimental band strength found for the NH$_2$ stretching modes of 2AO that appear around 3000 cm$^{-1}$ is 9.7 x 10$^{-17}$ cm/molecule. This value is close to the band strength measured by Brucato et al. (2006), for the equivalent mode in formamide (HCONH$_2$) ice. They give a band strength A= 1.35 x 10$^{-16}$ cm molec$^{-1}$ for ν$_1$ and ν$_2$ NH$_2$ modes in amorphous formamide at 20 K.

Figure 2 also shows the spectrum calculated with the methodology described in the previous section, for a crystalline solid. It can be seen that the simulation reproduces reasonably well the absorption features and relative intensities. The simulation provides also absolute band strengths for the different absorptions that have been presented in Table 1, together with those from the experiment. The theoretical band strengths are larger than the experimental ones, and differ by a factor that varies between 2.5 and 6, depending on the particular absorption. The theoretical simulation has guided the assignments of the different modes, which are given in the last column of Table 1.

*3.2. Low temperature 2-aminooxazole IR spectra.*

Figure 3 shows the infrared spectrum of 2AO as deposited at 20 K from the vapour phase, and after annealing to 180 K. An amorphous phase is formed when 2AO, molecules are deposited on a surface. At 20 K, the deposition is largely ballistic and gives rise to an amorphous solid. The deposited molecules do not have enough energy to reorganize into a crystalline network, this is reflected in the rounded profiles of the IR band*s* (bottom trace in Figure 3). Then, a transformation from an amorphous to a crystalline phase takes place when the ice is annealed from 160 K to 170 K. The spectrum of the polycrystalline sample formed in this way (middle trace in Figure 3) is slightly different from that of the crystalline 2AO powder embedded in a KBr pellet at room temperature (top trace in Figure 3). The deposit sublimates when the temperature is raised to 200 K.

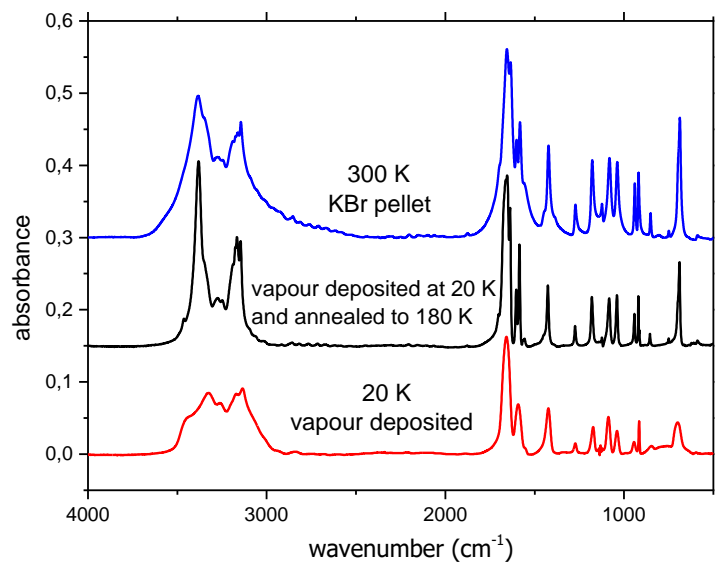

Figure 3. IR spectra of 2-aminooxazole vapour deposited at 20K (red) and annealed to 180 K (black) in comparison with the IR spectrum of a 2-aminooxazole KBr pellet (blue).

As can be seen in Figure 3, the amorphous phase obtained by vapour deposition presents broader features than the crystalline phase. Also some peaks are blended into single absorptions. The bands of both low temperature phases have been identified and assigned by comparison with the bands of the room temperature spectrum. Peak positions and wavenumber ranges of the bands of these two low temperature phases are listed in Table 2. The normalized band strengths of modes of the crystalline phase (named crystalline 180K) with respect to the intensity of the ν-$NH_2$ mode, and the relative band strengths of the amorphous phase with respect to the crystalline one, $A_{amorp}/A_{crysta180K}$, are given in the last two columns of Table 2.

Table 2. Observed IR absorption wavenumbers ($cm^{-1}$). Assignment of the spectrum of crystalline and amorphous 2-aminooxazole, and relative band strengths for the observed absorptions. Amorphous ice was grown by vapour deposition at 20 K, and annealed at 180 K to crystallize. The last column indicates the band strength ratio between these two ices. Symbols ν, β, δ, ρ and ω stand for stretch, deformation, bending, rocking and wagging vibrations respectively. Subindices a and s denote antisymmetric and symmetric modes.

| Peak crystalline ($cm^{-1}$) | Wavenumber range ($cm^{-1}$) Crystalline 180K | Peak amorphous ($cm^{-1}$) | Wavenumber range ($cm^{-1}$) amorphous | band assign. | $A_{crysta180K}$ / $A_{crysta180K}$ (ν -$NH_2$) | $A_{amorp}$ / $A_{crysta180K}$ |
|---|---|---|---|---|---|---|
| 3464, 3381, 3275, 3166, 3145 | 3660-2885 | 3429, 3327, 3264, 3157, 3135 | 3760-2887 | ν $NH_2$ | 1 | 1.04 |
| 1661,1653 1637,1603, 1586, 1561 | 1930-1473 | 1659,1592 1553 | 1840-1520 | δ $NH_2$ | 0.4095 | 1.22 |
| 1426 | 1473-1293 | 1423 | 1520-1363 | $\nu_s$OCO +$\nu_a$OCN | 0.0474 | 0.76 |
| 1274 | 1293-1220 | 1273 | 1306-1241 | β ring | 0.0079 | 1.30 |
| 1179 | 1213-1115 | 1173 | 1213-1147 | δ NCH | 0.0309 | 1.52 |
| 1083 | 1115-1063 | 1088 | 1117-1062 | ρ $NH_2$ | 0.0294 | 0.98 |
| 1040 | 1063-1007 | 1039 | 1062-1008 | ρ CH | 0.0225 | 1.27 |
| 943, 919 | 960-900 | 944, 916 | 970 - 902 | ω HCCH + ω $NH_2$ | 0.0198 | 1.23 |
| 854 | 870-837 | 845 | 883-824 | | 0.0033 | 0.69 |
| 690 | 740-630 | 701 | 733 - 656 | ρ HCCH | 0.0495 | 1.23 |

Figure 4 shows the spectrum of a mixture of 2AO with water (1:2) (2AO:$H_2O$), in comparison with the spectrum of the pure species. In the spectrum of the mixture, some of the bands of 2AO appear on top of the intense water absorptions at ≈ 1660 and 760 cm$^{-1}$. A slight high-frequency shift due to the presence of water can be appreciated.

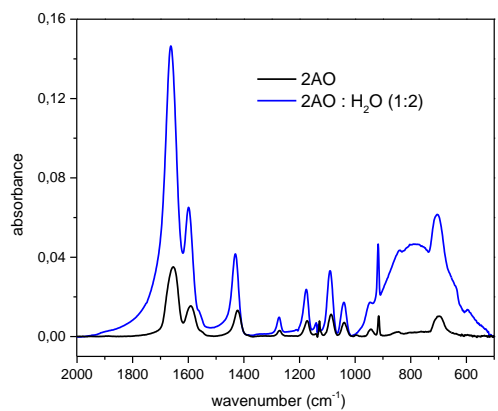

Figure 4. IR spectra of pure 2-aminooxazole (black) and of a mixture with water (blue) in a proportion 2AO : $H_2O$ (1:2). Both samples were deposited at 20 K

*3.3. Main products formed after processing with UV photons or 5 keV electrons.*

During the processing of 2-aminooxazole ices, pure or mixed with water, new infrared absorption peaks appear, corresponding to species formed upon destruction of the precursor. Figure 5 shows the spectra before and after energetic processing with UV photons or 5 keV electrons. In the spectral region between 2500 and 2000 cm$^{-1}$, 2AO has no IR absorptions and then, it is easy to recognize all the peaks corresponding to the new species. This is not the case for wavelengths below 1900 cm$^{-1}$, where there are plenty of absorption peaks corresponding to the original molecule, and some bands of newly formed species that overlap with absorptions of 2AO.

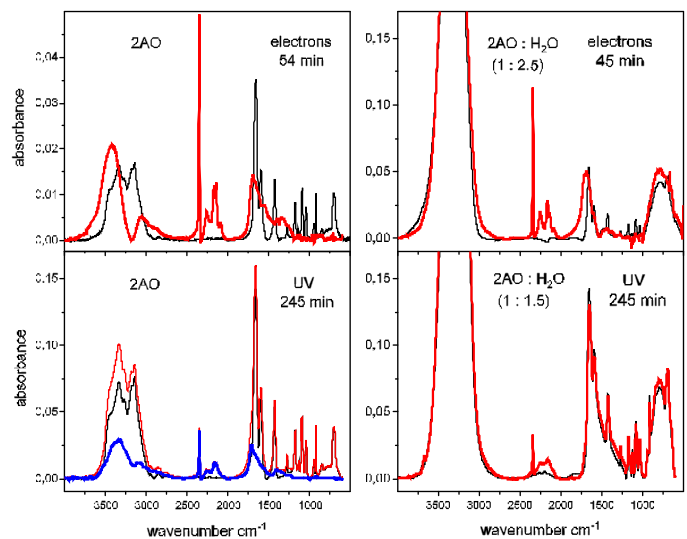

Figure 5. IR spectra of 2-aminooxazole (left panels) and 2AO:H$_2$O mixtures (right panels) deposited at 20 K. Black traces: initial spectra, before processing. Red traces: final spectra, after 54 min or 45 min processing with 5 keV electrons (top panels), or 245 min processing with UV photons (bottom panels). In all cases the contribution of residual water ice deposited during processing has been subtracted. Left bottom panel blue trace: final spectrum after subtraction of the contributions of unprocessed 2-aminooxazole, estimated to be 0.8 x initial spectrum (see text).

In the bombardment with 5 keV electrons, the penetration depth was larger than the ice thickness and all the original 2AO was destroyed at the end of the processing time. Therefore, no treatment, apart from a baseline subtraction and the removal of the ASW background water layer grown during deposition, was needed for the spectra in the top panels. The penetration depth of electrons in the samples was estimated with the CASINO code (Drouin et al. 2007, Drouin 2011). Simulations, containing both the sample and the Si substrate, showed that the amount of secondary electrons backscattered from the substrate into the ice is negligible. In the case of UV processing, a large fraction of the precursor remains at the end of the experiments. The amount of unprocessed 2AO was found to be ≈ 80% by observing the decay of the bands below 1200 cm$^{-1}$. In this spectral region there are no appreciable product absorptions and the decrease in the intensity of the bands reflects the destruction of the precursor. The spectrum of unprocessed 2AO has been removed from the final spectrum presented in the lower left panel of Figure 5 (red curve), in order to facilitate the visualization of absorption bands of the newly formed products (blue curve). However, in this case, the final shape of the spectra could be slightly distorted (this problem does not affect the 2500-2000 cm$^{-1}$ region). Note the overall similarity between the blue spectrum in the lower left panel (products from UV processing) and the red spectrum in the upper left panel (products from electron processing).

The main absorption bands in the IR spectra of the products shown in Figure 5 are listed in Table 3 with suggested assignments. Although the various spectra of the products exhibit some dissimilarities, we have concentrated on the main common features and do not discuss these smaller differences, that may be affected by the just mentioned subtraction of the signals of water and unprocessed 2AO. A mixture of different compounds is formed after energetic processing of the ice samples. In general, the recognition of specific molecules is not possible with IR spectroscopy alone, but the analysis of the spectra allows the identification of the main functional groups in the reaction products.

Broad absorption bands in the 3600-2600 $cm^{-1}$ range are observed in the spectra of the processed samples of pure 2AO ice with maxima at about 3400 $cm^{-1}$. They are likely due to NH stretch vibrations of amines or amides. In the sample processed with electrons, a smaller band with a maximum at 3050 $cm^{-1}$ and a shoulder at 2875 $cm^{-1}$ is also seen, whereas in the UV irradiated deposit there is just one band with a secondary maximum at 3102 $cm^{-1}$ and a shoulder at 2875 $cm^{-1}$. The absorptions below 3100 $cm^{-1}$ have probably a high contribution of $CH_x$ stretch vibrations. In the mixtures with water ice, the intense OH band of $H_2O$ obscures the IR absorption of the products in this region.

A narrow absorption peak, due to $CO_2$, is found at 2343 $cm^{-1}$ in all the processed samples. This peak is much more intense in the deposits processed with high energy electrons than in those irradiated with UV photons. A structured absorption between 2260 and 2000 $cm^{-1}$, with four blended peaks, is also seen in all the samples. The first peak, at 2255 $cm^{-1}$, is attributed to asymmetric stretch vibrations of cyanate or isocyanate functional groups with possible contributions from stretching vibrations of CN triple bonds in nitriles. The peak at 2164 $cm^{-1}$ is assigned to the $OCN^-$ anion, which is usually found in the energetic processing of ices containing NCO subunits, and the one at 2137 $cm^{-1}$, to CO. Finally, the peak at 2085 $cm^{-1}$ is attributed to the CN stretching mode of HCN or of the $CN^-$ ion; this peak is rather a shoulder in the experiments with UV irradiation. CO and $CO_2$ evaporate when the processed samples are heated to 200 K, as evidenced by the disappearance of the peaks at 2343 and 2137 $cm^{-1}$ (see Figure 6). The peaks at 2255 and 2085 $cm^{-1}$ also decrease markedly or vanish altogether upon warming, suggesting that they are associated with volatile species like HNCO or HCN respectively. In contrast, the peak at 2164 $cm^{-1}$ remains when the samples are heated, indicating that $OCN^-$ is linked in some non-volatile ionic structures. A peak at about 2220 $cm^{-1}$, likely due to CN stretch vibrations of nitriles, is unburied after the disappearance of the 2255 $cm^{-1}$ maximum.

Table 3 IR absorptions in the IR spectra of the processed samples. Symbols $\nu$ and $\delta$ stand for stretch and bending vibrations respectively. Subindex a denote antisymmetric modes.

| Absorptions ($cm^{-1}$) | Assignment |
|---|---|
| 3600-2600<br>  Maximum:<br>  3400<br>  Secondary maxima or shoulders:<br>  3100-3050 | <br><br>$\nu\ NH_2$, $\nu\ NH$; amines, amides<br><br>$\nu\ NH$ ; amides, $\nu\ CH$ |

| 2876 | $\nu$ CH$_x$ |
|---|---|
| 2343 | $\nu_a$ CO$_2$ |
| 2280-2000 Peaks (blended) | |
| 2255 | $\nu_a$ -ONC, -NCO; cyanates, isocyanates |
| | $\nu$ -C≡N ; nitriles |
| 2164 | $\nu_a$ OCN$^-$ |
| 2137 | $\nu_a$ CO |
| 2085 | $\nu$ HCN, CN$^-$ |
| 2000-1200 Maximum: | |
| 1710-1690 | $\nu$ C=O carbonyl, amide I band |
| Secondary maxima or shoulders: | |
| 1630-1620 | |
| 1550 | $\delta$ NH |
| 1400 | Amide II band (o.p. $\delta$ NH + $\nu$ CN) |
| 1350 | $\delta$ CH$_x$ |
| 1250 | $\nu$ C-N, amide band III (i. p. $\nu$ CN + $\delta$ NH) |
| | $\nu$ C-O-C |

The spectral interval between 2000 and 1200 cm$^{-1}$ is characterized by a continuous broad absorption feature with a marked maximum at 1710-1690 cm$^{-1}$ and with some secondary maxima or shoulders extending towards the lower wavenumber range. IR absorptions in this region are dominated by stretching vibrations of the carbonyl C=O group, and by NH bending motions. Absorptions at 1550 and 1350 cm$^{-1}$ are suggestive of amide II and III bands, which correspond to out-of-phase and in phase NH bending plus C-N stretching vibrations of peptide bonds respectively (Barth & Zscherp 2002; Socrates 2001). The amide I band, typically at 1690-1620 cm$^{-1}$ and mostly due to CO stretching vibrations, can contribute largely to the main maximum and, specially, to its lower wavemumber wing. The solid material formed at 20 K is modified upon heating, as shown in Figure 6 by the band profile changes in the 2000-1200 cm$^{-1}$ spectral interval, but a non-volatile residue still remains when the temperature is raised to 200 K.

Figure 6 shows the effect of warming on the spectra of the products obtained at the end of the processing experiments.

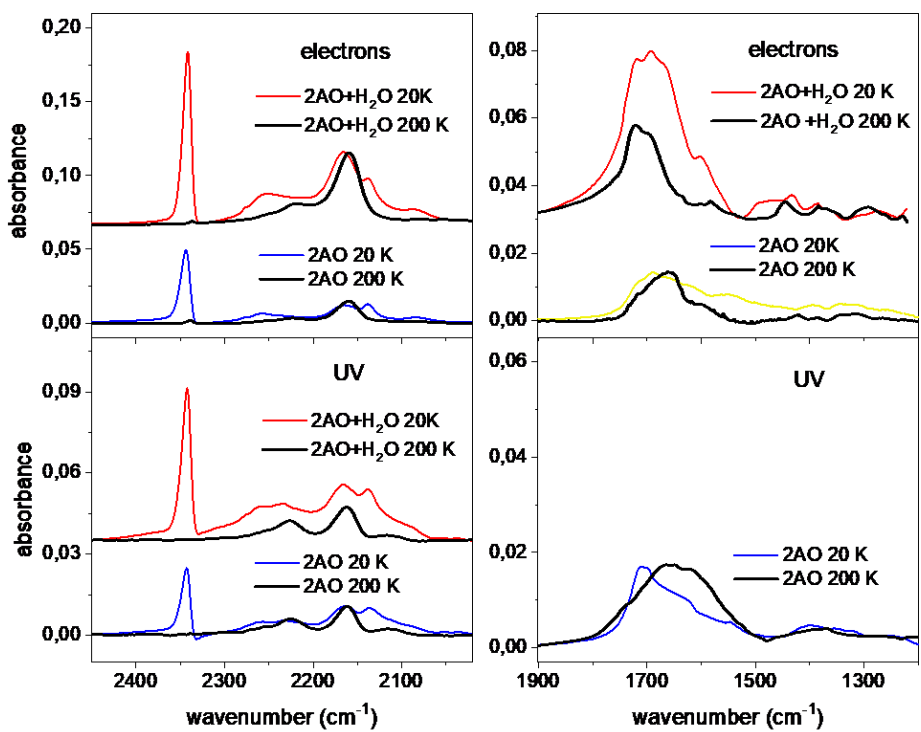

Figure 6. IR spectra of products of processing of 2AO, both for the pure species and for water mixtures. Red or blue traces: 20 K. The spectra have been displaced vertically to aid the eye. Black traces: after warming to 200 K. Right bottom panel: only the spectra of pure 2AO is shown. The 2AO:H$_2$O mixture proportions and processing times at 20 K are the same as in Figure 5.

The present experiments do not provide information on the details of the chemistry induced by electron bombardment or photon irradiation on the 2AO deposits. Theoretical studies on proton-induced charge transfer (Bacchus Montabonel 2015, Calvo et al. 2016) and on UV photochemistry (Szabla et al. 2013, 2015) have been reported for 2AO molecules, both bare and in clusters with water molecules. The calculations of Szabla et al. (2013, 2015) on the effects of UV excitation demonstrated the relevance of H atom photodetachment from the NH$_2$ group. These authors also identified an important ring opening channel at the CO bond. The photodegradation of 2AO in aqueous solution with photons in the 215-285 nm (5.67-4.35 eV) range has recently been investigated by Todd et al. (2019), but, as far as we know, no studies of the photoproducts have been reported thus far

3.3. Destruction cross sections and Half-life Doses.

A list of the processing experiments performed in this work is given in Table 4. The column densities of 2-aminooxazole and ASW have been calculated with A$_{2OA}$(1423 cm$^{-1}$)= 6.6 x 10$^{-18}$ cm molec$^{-1}$ (Table 1) and A$_{ASW}$(3000 cm$^{-1}$)= 2.9 x 10$^{-16}$ cm molec$^{-1}$, (Mastrapa et al., 2009), respectively, and are given in the first two columns of the table. The layer thickness for pure 2-aminooxazole ices has been calculated assuming a density of 1.24 g cm$^{-3}$. For mixtures, an average density has been estimated considering 0.65 g cm$^{-3}$ the density of vapour deposited ASW at 20 K (Dohnálek, 2003). Destruction cross sections are derived from decay data after 15-30 min irradiation and during that time background ice grows just by 1.5-3 nm as compared with sample thicknesses in the 60-550 nm range.

Table 4. Processing experiments at 20 K. The molecular ratio (2AO:H2O) is given in the first column and the column densities of the two species in the next two columns. The fourth column contains the estimated sample thickness (L). Energies and fluxes of the photons and electrons impinging on the ice surface are given in the last two columns. Column densities for H$_2$O and 2AO have an uncertainty of the order of 25%, due to the inaccuracy of the band strength of these species. Pure 2AO ice layer thickness has similar uncertainties, 25%, for similar reasons. The estimated uncertainty in L for mixtures is about 30%, due to the error added for considering an average density.

| Sample | N$_{2AO}$ (molec cm$^{-2}$) | N$_{ASW}$ (molec cm$^{-2}$) | L (nm) | Processing agent | φ$_0$ (cm$^{-2}$ s$^{-1}$) |
|---|---|---|---|---|---|
| UV1: 2AO | 4.1 x 10$^{17}$ | | 482 | UV (10.3-6.9 eV) | 7.5 × 10$^{13}$ |
| UV2: 2AO:H$_2$O (1:1.5) | 3.6 x 10$^{17}$ | 5.4 x 10$^{17}$ | 552 | UV (10.3-6.9 eV) | 7.5 × 10$^{13}$ |
| UV3: 2AO | 5.2 × 10$^{16}$ | | 61 | UV (10.3-6.9 eV) | 7.5 × 10$^{13}$ |
| E1: 2AO | 1.4 x 10$^{17}$ | | 160 | Electrons, 5 keV | 4 × 10$^{12}$ |
| E2: 2AO:H$_2$O (1:3.5) | 1.2 x 10$^{17}$ | 4.2 x 10$^{17}$ | 289 | Electrons, 5 keV | 4 × 10$^{12}$ |

Destruction cross sections for 2-aminooxazole under UV irradiation or electron bombardment are estimated from the decay of the peak at 1423 cm$^{-1}$, which was assigned to the OCN asymmetric stretching in the ring (see Table 1). This band appears quite isolated in the spectrum, and is not disturbed either by bands of the products or by water bands (a contamination always present in the high vacuum setup). Assuming first order irreversible kinetics, the evolution of this band is given by:

$$\ln\left(\frac{I_F}{I_0}\right) = -\sigma_{des} F \qquad (1)$$

Where $I_F$ is the band intensity (absorbance) at a particular fluence, $I_0$ is the band intensity at the beginning of the experiment, $\sigma_{des}$ is the destruccion cross section (cm$^2$ molecule$^{-1}$ photon$^{-1}$ or electron$^{-1}$), F the fluence (photons or electrons cm$^{-2}$).

*3.3.1. UV destruction cross section*.

VUV radiation is substantially attenuated by ices. As discussed by Öberg (2016), there are some discrepancies in the experimental literature on the UV irradiation of ices most probably arising from the different thicknesses of the samples used by various groups, which were generally treated as optically thin layers. In the present experiments two thick (≈ 500 nm) layers **(UV1 and UV2** in Table 4) and a thin (60 nm) layer **(UV3** in Table 4) were used. The decay of the 1423 cm$^{-1}$ band as a function of irradiation time is displayed in the upper panel of Figure 7. Note the expected faster decline of the band in the thinner layer.

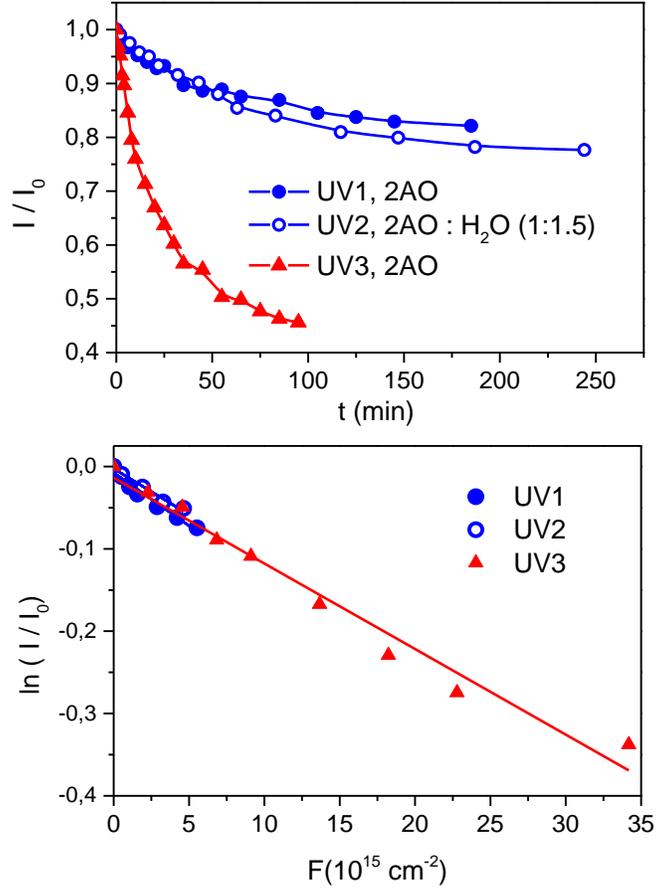

Figure 7. Upper panel: normalized integrated intensity, of the 1423 cm$^{-1}$ band of 2-aminooxazole (OCN asymmetric stretch in the ring) vs UV processing time at 20 K. UV1-UV3 correspond to the samples listed in Table 4. Lower panel: Logarithmic representation of the exponential decay of the band vs photon fluence observed at the beginning of UV processing (see text). The straight lines correspond to a fit of the data with equation (1). For the three samples, the data represented in the lower panel correspond to approximately 15 minutes of irradiation.

To derive destruction cross sections from decay measurements on thick ice layers, one cannot employ the value of the UV photon flux impinging on the surface, $\phi_0$. Instead, the average flux, $\phi_{av}$, inside the ice sample must be used. It can be calculated using Beer´s Law:

$$\phi_{av} = \frac{1}{N_0} \int_0^{N_0} \phi_0 e^{-\sigma_a N} dN = -\frac{\phi_0}{N_0 \sigma_a} (e^{-\sigma_a N_0} - 1) \qquad (2)$$

where $N_0$ is the initial (total) column density, and $\sigma_a$ is the absorption cross section of the solid. For binary mixtures, $N_0 = N_1 + N_2$ and $\sigma_a$ is the absorption cross section of the ice mixture, which is in general not known. We have assumed $\sigma_a \approx f_1\sigma_a(1) + f_2\sigma_2(2)$, where indexes 1 and 2 correspond to the two mixture components, and N, f and $\sigma_a$ stand for column density, relative fraction, and absorption cross section respectively.

UV absorption cross sections for ices of the most common small molecules of astrophysical interest have been recently measured by Cruz-Diaz et al. (2014a, b) in the ≈ 120-180 nm range, but for other molecules they are mostly unknown. The works of these authors showed that, despite the differences in the profile of the gas-phase and ice spectra, the average absolute values of the cross sections are often of the same order over the 120-180 nm range. Figure 8 shows the UV absorption spectrum of room temperature of solid 2AO measured in this work. An absorption continuum is found in the 120-250 nm (10.3-4.9 eV) interval, with a maximum at 158 nm (7.8 eV), and broad secondary maxima at larger wavelengths. About half of the observed absorption takes place in the 120-180 nm range, which corresponds to the output of our $D_2$ lamp.

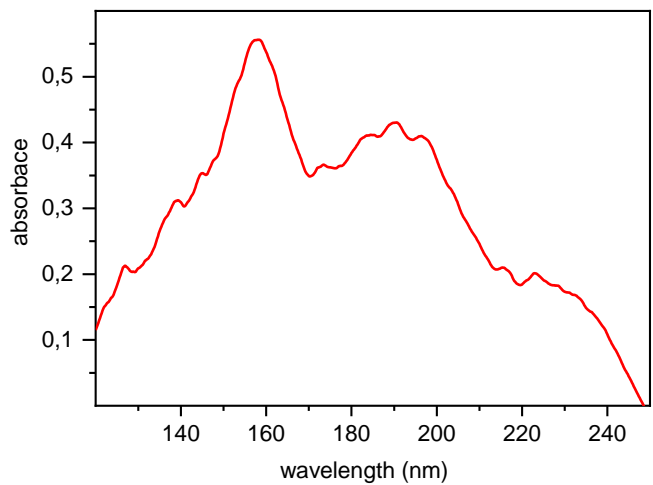

Figure 8 UV absorption spectrum of solid 2AO at room temperature

Unfortunately, absolute values of the absorption cross sections could not be derived from the measurements and, as far as we know, they have not been reported in the literature neither for solid 2AO nor for the gas-phase molecule. There are some cross section measurements for UV absorption by gas-phase related heterocycles like oxazole (Palmer et al. 2007) and isoxazole (Walker et al. 2004). In the 120-250 nm range, these molecules exhibit also continuous absorption, and their spectra, largely dominated by ππ* transitions, have a qualitatively similar profile to that of solid 2AO, with maxima close to 155 nm and broad secondary maxima at 200 nm with absorption extending down to 250 nm. Over the 120-180 nm interval the average absorption cross-section for oxazole and isoxazole is ≈ 3 x $10^{-17}$ $cm^2$. We have assumed a value of 3 ± 0.5 x $10^{-17}$ $cm^2$ for 2AO given the lack of a more precise estimate. The absorption cross sections and average fluxes inside the ice layers can be found in Table 5. The relevant fluences for the irradiation of our samples are obtained as F= $\phi_{av}$ Δt.

The levelling-off of the decay curves observed in Figure 7 is a common phenomenon observed in the photolysis of ices. This behaviour deviates from the exponential decay implied by equation 1, especially for thick layers, and it is usually attributed either to reformation of the original precursor through backward reactions, or to changes in the optical properties of the sample due to the appearance of the new reaction products and also to the deposition of an outer layer of background water molecules, which is unavoidable in long measurements under high vacuum ($P_b \approx 10^{-8}$ mbar in our chamber). Consequently, the validity of equation (1) is restricted to the beginning of the irradiation process, and it is customary to derive destruction cross sections just from the initial data, corresponding to the exponential decay (see, for instance, Gerakines et al. 1996, Öberg et al. 2009). The decay of the band vs fluence in this initial range is shown in the lower panel of Figure 7. From a fit of these data to equation (1) we have obtained the destruction cross sections listed in Table 5.

The $\sigma_{des}$ values are in the 0.8-1.0 × $10^{-17}$ cm$^2$ range The good agreement between the destruction cross sections of 2AO derived from layers UV1 ($\approx$ 550 nm) and UV3 ($\approx$ 60 nm), is reassuring and lends support to the consistency of the method employed. Given the high absorption cross section of 2AO, the attenuation of UV radiation by the samples is high and cannot be ignored even for the thinner layer. The average photon flux inside sample **UV3** is about one half of that of the impinging radiation. Within the large experimental uncertainty, ices of pure 2AO and of 2AO:H$_2$O decay at a very similar rate. Water molecules modify appreciably the absorption cross section of the sample and thus the average photon flux within it, but otherwise they do not seem to play a role in the UV destruction of 2AO.

Table 5 Estimated UV (120-180nm, 6.9-10.3 eV) absorption cross sections, $\sigma_a$, destruction cross sections, $\sigma_{des}$, and average photon fluxes ($\phi_{av}$, equation 2) inside the ice samples of pure 2-aminooxazole and of mixtures of 2-aminooxazole with water. The corresponding column densities are listed in Table 4. For the calculations of $\sigma_a$ in the mixtures, we have assumed an average $\sigma_a$(H$_2$O) $\approx$ 3 × $10^{-18}$ cm$^{-2}$ for water-ice in this spectral range (Cruz-Diaz et al. 2014a). The errors are qualitative estimates, since for $\sigma_a$ there are no direct measurements, and the procedure to determine $\sigma_{des}$ involves quite crude approximations.

| sample | $\phi_{av}$ (photons cm$^{-2}$ s$^{-1}$) | $\sigma_a$ (cm$^2$) | $\sigma_{des}$ (cm$^2$) |
|---|---|---|---|
| **UV1** 2AO | 6.1 × $10^{12}$ | 3.0 ± 0.5 × $10^{-17}$ | 0.9 ± 0.5 × $10^{-17}$ |
| **UV2** 2AO:H$_2$O (1:1.5) | 6.0 × $10^{12}$ | 1.4 ± 0.5 × $10^{-17}$ | 0.8 ± 0.5 × $10^{-17}$ |
| **UV3** 2AO | 3.8 × $10^{13}$ | 3.0 ± 0.5 × $10^{-17}$ | 1 ± 0.5 × $10^{-17}$ |

The cross sections found in the three experiments (UV1-UV3) lie within the mutual experimental error. Values of 9.5 ± 5 × $10^{-18}$ cm$^2$ (average of the UV1 and UV3 samples) and 8 ± 0.5 × $10^{-18}$ cm$^2$ were obtained for the pure and mixed 2AO samples respectively. The results of this work are compared in Table 6 with destruction cross sections for various organic molecules under different conditions of interest for prebiotic chemistry, ranging from interstellar ice analogues to aqueous solutions. The UV photon sources in these experiments were mostly H$_2$(D$_2$) lamps, that provide VUV photons useful for the simulation of the interstellar UV field, and Xe lamps, more adequate for the generation of longer wavelength photons, abundant in the solar radiation. The photon energy distributions provided by the lamps depend on the conditions of operation and thus, for a given lamp type, they may vary between the different experiments. Todd and coworkers (Todd et al., 2019) investigated the UV photostability of three related molecules in the 2AO family (2 aminooxazole, aminoimidazole, and aminotiazole) at larger wavelengths. These species were dissolved in water, at a 0.1 mM concentration, and processed with UV radiation from a Xe lamp coupled with a diffraction grating, which allows wavelength selection between 215 and 285 nm. The cross section they found for 2AO, 2.1 x $10^{-18}$ cm$^2$, is lower than that of the present work for photolysis with smaller wavelengths.

Table 6. 2-aminooxazole destruction cross sections obtained in this work compared with literature values for other organic molecules in condensed phases.

| Species | T (K) | UV source λ(nm) | $\sigma_{des}$(cm$^2$) | Reference |
|---|---|---|---|---|
| 2AO ice, pure | 20 | D$_2$ lamp 120-180, peak 160 | 9.5 ± 0.5 × 10$^{-18}$ | This work |
| 2AO : H2O (≈1:1.5) | 20 | D$_2$ lamp 120-180, peak 160 | 8 ± 0.5 × 10$^{-18}$ | |
| 2AO (0.1mM) | 296 | Xe lamp 215 | 2.1 x 10$^{-18}$ (a) | Todd et al. 2019 |
| 2AO (0.1mM) | 296 | Xe lamp 285 | 1.9 x 10$^{-19}$ (a) | |
| Methylisocyanate | 20 | D$_2$ lamp 120-180, peak 160 | 3.7 x 10$^{-18}$ (b) | Maté et al. 2018 |
| Methylisocyanate(4%):H2O | 20 | D$_2$ lamp 120-180, peak 160 | 2.4 x 10$^{-18}$ (b) | |
| Methanol (CH3OH) | 20 | H$_2$ lamp 115-170, peak 122 | 1.6 x 10$^{-18}$ | Gerakines et al. 1996 |
| Methanol (CH$_3$OH) | 20 | H$_2$ lamp 100-200, peak 162 | 0.5 x 10$^{-18}$ | Cottin et al. 2003 |
| Methanol (CH3OH) | 20 | H$_2$ lamp 115-170, peak 122 | 2.6 x 10$^{-18}$ | Öberg et al. 2009 |
| Formaldehyde (H2CO) | 13 | H$_2$ lamp 115-170, peak 122 | 6.2 x 10$^{-18}$ | Gerakines et al. 1996 |
| Formaldehyde (H2CO) | 13 | H$_2$ lamp peak 122 | 7.5 x 10$^{-18}$ | Butscher et al. 2016 |
| Formamide(HCONH2) on minerals 63 K | 63 | Xe lamp ≈ 260 | 3.7 x 10$^{-19}$- 1 x 10$^{-20}$ | Corazzi et al. 2020 |
| Glycoaldehyde (HCOCH2OH) | 20 | Xe lamp 285 | 1.1 x 10$^{-20}$ | Puletti, thesis, 2014 |
| Glycine+H2O | 12 | H$_2$ lamp 110-185, peak 122 | 6.9 x 10$^{-19}$ (c) | Ehrenfreund et al. 2001 |
| Alanine+H2O | 12 | H$_2$ lamp 110-185, peak 122 | 8.6 x 10$^{-19}$ (c) | Ehrenfreund et al. 2001 |
| Glycine | 18 | Xe lamp, 147 | 3.5 x 10$^{-19}$ (c) | Johnson2012 |
| Glycine | 18 | Hg lamp, 254 | 4.8 x $10^{-20}$ (c) | Johnson2012 |

(a) These values were obtained by multiplying the destruction rates (min$^{-1}$) given in the supplementary material of Todd et al. (2019) by the normalized UV flux. (b) These values were obtained by Maté et al. (2018) under the assumption of optically thin films. Using equation (2) of this work with an absorption cross section of $\sigma_a \approx 2 \times 10^{-17}$ cm$^2$ (Tokue et al. 1986) to correct for the actual film thickness, the destruction cross sections transform to 5.6 × $10^{-18}$ (pure) and 6.3 × $10^{-18}$ cm$^2$ (mixture). The astrophysical implications discussed in Maté et al (2018) remain unchanged. (c) These values were obtained from the half-lives and fluxes given in the publications using equation (3).

Overall, the UV destruction cross sections found in this work for 2AO ice are larger than all the rest, reflecting the low photostability of this molecule also in the VUV range. These large cross sections are not entirely surprising and are associated with the strong VUV absorption observed in gas-phase N heterocycles (Palmer et al. 2007, Walker et al 2004). Hydrogen lamps were employed by other authors to obtain destruction cross sections of methanol (CH$_3$OH) (Gerakines et al. 1996, Oberg et al. 2009) and formaldehyde (H$_2$CO) (Gerakines et al., 1996, Butscher et al. 2016). The cross sections found lie mostly in the $10^{-18}$ cm$^2$ range. A larger wavelength, UV radiation, at 285 nm, provided by a Xe lamp coupled to a monocromator, was used by Puletti (2014) to process glycoaldehyde (HCOCH$_2$OH) ice. A recent work by Corazzi et al. (Corazzi et al., 2020) investigates the photo destruction of formamide (HCONH$_2$) ice deposited at 63 K on top of different minerals, with a Xenon lamp in the 230- 300 nm range. For the larger wavelengths (≈ 260-280 nm) employed in the last two works, the destruction cross sections found are in the $10^{-20}$ cm$^2$ range. The UV photostability of glycine at low temperature, has been investigated over a larger wavelength range, from 147 nm to 254 nm UV (Johnson et al., 2012) and the cross sections found vary from 3.5 x $10^{-19}$ to 4.8 × $10^{-20}$ cm$^2$. Ehrenfreund et al. (2001) investigated the stability of glycine:water ices with a microwave discharge H$_2$ lamp and found a destruction cross section of 6.9 × $10^{-19}$ cm$^2$.

*3.3.2. 5 keV electron destruction cross section.*

The experiments of electronic processing, E1 and E2 in Table 4, will be analysed next. In this case, the thicknesses of the irradiated 2-aminooxazole layers are well below the 500 nm penetration depth of the electrons estimated with the CASINO code (Drouin et al. 2007, Drouin 2011) for 5 keV electrons impinging on 2AO. Using this program, the linear energy transfer for 5 keV on a 160 nm layer of 2AO with 1.2 g/cm$^3$ density, has been estimated to be 6.1 keV/µm. The decay of the 1423 cm$^{-1}$ band versus processing time is presented in the upper panel of Figure 9.

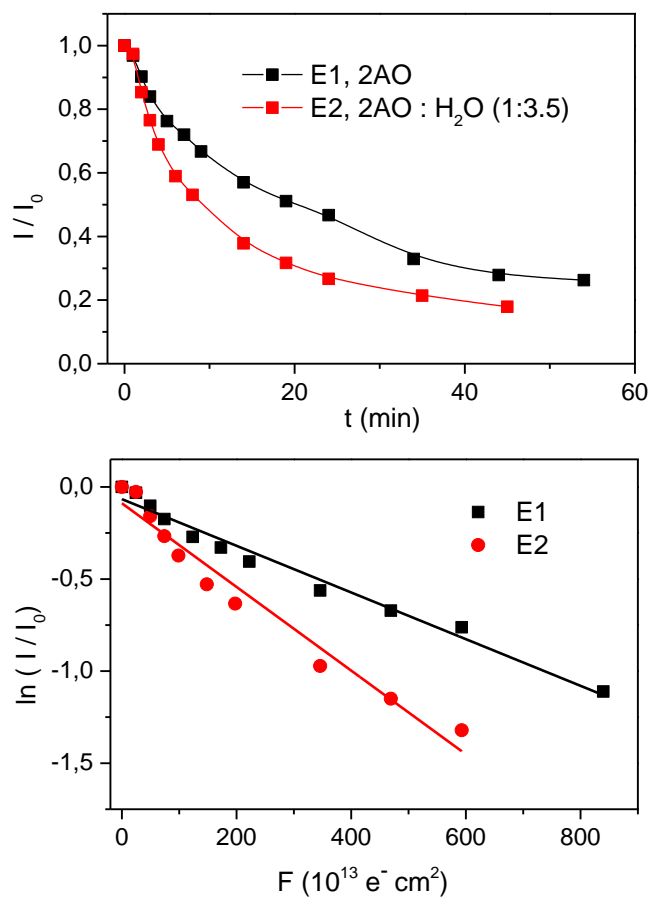

Figure 9. Upper panel: normalized decays of the 2-aminooxazole 1420 cm$^{-1}$ band versus processing time with 5 keV electrons of samples E1 (pure) and E2 (2AO(22%)/H$_2$O mixture). Lower panel: Logarithmic representation of the data in the upper panel vs electron fluence in the region of exponential decay (see text). The lines are fit of the data to equation (1). The points represented in the lower panel correspond to irradiation times of 25 min (E2) and 35 min (E1)

A logarithmic representation of the normalized intensity decay vs electron fluence gives an approximate linear behaviour that can be fitted to equation (1) to obtain destruction cross sections. The fits are displayed in the lower panel of Figure 9 and the 5 keV electron destruction cross sections obtained from them are listed in Table 6, where they are also compared with previous literature results for other complex organic molecules (COMs) of astrophysical relevance.

Bombardment by high energy electrons induces a cascade of secondary electrons and bremsstrahlung photons similar to that produced by cosmic rays. Cosmic rays are largely made by protons in the MeV energy range and, as noted by Kaiser et al. (2013), the interaction of electrons of a few keV with a solid sample is characterized by a linear energy transfer (LET) of a similar order as that of protons in the MeV range. For a given substance, the destruction cross section is roughly proportional to the linear energy transfer, irrespective of the type of projectile, as exemplified in the recent work of Da Costa et al. (2020) on the radiolysis of valine by different ions. LET values for various COMs bombarded with MeV protons or keV electrons are also given in Table 6. Taking all the cases considered they vary within less than a factor 5.

Table 7. 5 KeV destruction cross sections, LET and half-life doses for 2-aminooxazole obtained in this work compared with previous literature data for other astrophysical relevant species.

| Experiments | T(K) | $\sigma_{des}$ (cm$^2$) | projectile | LET keV/μm | $D_{1/2}$ (eV molec$^{-1}$) | Ref |
|---|---|---|---|---|---|---|
| E1, pure 2AO | 20 | 1.3 x 10$^{-16}$ | 5keV e- | 6.1 | 37 | This work |
| E2, (22% 2AO in ASW) | 20 | 2.3 x 10$^{-16}$ | 5keV e- | 5.9 | 43 | This work |
| Methylisocyanate | 20 | 4.5 x 10$^{-16}$ | 5keV e- | 6.3 | 9 ± 2 | Maté et al. 2018 |
| Methylisocyanate (5% in ASW) | 20 | 1.6 x 10$^{-15}$ | 5keV e- | 3.8 | 18 ± 5 | Maté et al. 2018 |
| Glycine | 20 | 9.7 x 10$^{-16}$ | 2keV e- | 22 | 12 ± 2 | Maté et al. 2015 |
| Glycine | 15 | 2.2 x 10$^{-15}$ | 0.8 MeV H+ | 45(a) | 13 ± 1.4 | Gerakines et al. 2012 |
| Alanine | 15 | 3.3x 10$^{-15}$ | 0.8 MeV H+ | 41(a) | 1 ± 0.4 | Gerakines et al. 2012 |
| Phenylalanine | 15 | 2.8 x 10$^{-15}$ | 0.8 MeV H+ | 37(a) | 25 ± 3.6 | Gerakines et al. 2012 |
| Valine | 300 | 2.6 x 10$^{-15}$ | 1.5 MeV H+ | 26 | 26(b) | Da Costa et al. 2020 |

(a) These values were obtained by multiplying the stopping power values by the densities reported in Table 1 of Gerakines et al. 2012 (b) The half-life dose of valine is estimated by assuming $D_{1/2}=\ln(2)LET/(\sigma_{des} n)$, where n is the molecular density of valine. This expression for $D_{1/2}$ is readily derived from equation (4) (see Section 3.3.3).

In the electron processing experiments of the present work, the presence of water has an appreciable effect on the destruction rate. The cross section obtained for the mixture is almost a factor two larger than that for the pure species. A similar tendency was observed for methyl isocyanate, where a 5% dilution of the molecule in water ice was found to increase the destruction cross section by a factor of about four (Maté et al. 2018).

*3.3.3 Half-life doses.*

A convenient way to measure the destruction efficiency of UV photons or electrons, is the half-life dose, $D_{1/2}$. It is defined as the amount of energy needed to reduce the initial number of molecules of a sample by one half and it is used to estimate the survival probabilities in different astronomical environments.

The half-life fluence, i.e., the fluence needed to decrease the initial concentration by one half, can be derived from equation (1) as:

$$F_{1/2} = \ln(2)/\sigma_{des} \quad (3)$$

The half-life dose can then be calculated as:

$$D_{1/2} = \frac{F_{1/2} f_E E}{N_{2AO}} = \frac{\ln(2) f_E E}{\sigma_{des} N_{2AO}} \quad (4)$$

where $f_E$ is the fraction of energy deposited by the photons or electrons in the sample, E is the photon or electron energy, and $N_{2AO}$ the column density of 2AO at the beginning of the experiment.

Equation (4) is valid for optically thin layers, where the whole sample is effectively processed. If the layer is too thick, the majority of the sample remains unprocessed and unrealistically small values are derived. Therefore, for the thick ice layers used in the UV irradiation experiments of this work, we must take the optically thin limit in expression (4).

The fraction of UV energy absorbed by an ice layer of column density N can be expressed as:

$$f_E = 1 - \frac{\phi_N}{\phi_0} = 1 - e^{-\sigma_a N} \quad (5)$$

where $\phi_N$ is the photon flux transmitted through N. Taking now the thin layer limit, equation (4) transforms into:

$$D_{1/2} = \lim_{N \to 0} \frac{\ln(2)(1-e^{-\sigma_a N}) E}{\sigma_{des} N} = \ln(2) E \frac{\sigma_a}{\sigma_{des}} \quad (6)$$

For a binary mixture, the half-life dose of species *i* is given by:

$$D_{1/2}(i) = \lim_{N \to 0} \frac{\ln(2)(1-e^{-\sigma_a N}) E}{\sigma_{des}(i) f_i N} = \frac{\ln(2) E}{f_i} \frac{\sigma_a}{\sigma_{des}(i)} \quad (7)$$

where $f_i$ and $\sigma_{des}(i)$ are the relative fraction and destruction cross section of component *i* in the mixture respectively. The product $f_i N = N_i$ is the column density of component *i*, and $\sigma_a$, is the absorption cross section of the ice, which depends on the mixture proportion. In the absence of more precise data, it can be approximated by $\sigma_a \approx f_1 \sigma_a(1) + f_2 \sigma_a(2)$, as indicated in the previous section. The half-life doses calculated with equations (6) and (7) for the conditions of the laboratory measurements are listed in Table 8. The average energy of the photons from the UV lamp is 8.1 eV (Maté et al. 2018) and the $\sigma_a$ and $\sigma_{des}$ values are given in Table 5.

As discussed elsewhere (Johnson & Quickenden 1997, Maté et al. 2018), the main effect of water in binary ices with relative large molecules is to provide a shield by absorbing part of the UV radiation. The nascent photofragments of water, H and OH, recombine mostly to $H_2O$ again and do not participate in the destruction of 2AO. Thus, the presence of water has little effect on the cross section, but increases the half-life dose.

Table 8. Half-life doses for ices of pure 2-aminooxazole and of 2-aminooxazole with water irradiated with a $D_2$ lamp (120-180nm) with an average photon energy of 8.1 eV (Maté 2018).

| Sample | $f_{2AO}$ | $D_{1/2}$ (eV molecule$^{-1}$) |
|---|---|---|
| UV1: 2AO | 1 | 18.1 |
| UV2: 2AO:H$_2$O (1:1.5) | 0.40 | 24 |
| UV3: 2AO | 1 | 16.2 |

In the bombardment with high energy electrons, the mechanism is different. Instead of the gradual attenuation of the photon beam as it travels through the ice, the cascade of secondary electrons induced by the high energy particles leads to a more homogeneous energy release along the high-energy electron beam path. If the penetration depth of the electrons is larger than the width of the ice layers, as it is the case in the experiments of this work, an approximately linear energy transfer can be assumed. The fraction of energy absorbed, $f_E$, is calculated with the CASINO code (Drouin et al. 2007, Drouin 2011), and is related with the values of the linear energy transfer given in Table 7. Specifically, the values of $f_E$ for pure 2AO (sample E1) and 2AO (22%) + H$_2$O (sample E2) are 0.20 and 0.34 respectively. The corresponding half-life doses are also given in Table 7, where they are compared with those of methyl isocyanate and various amino acids. The half-life doses for 2AO are the largest of the table and indicate that 2AO ice is somewhat more stable than the rest of the species listed against the bombardment by energetic particles.

4. Astrophysical implications.

As mentioned in the introduction, 2-aminooxazole is involved in the ribonucleotide synthesis pathway advanced by Powner et al. (Powner et al., 2009). These authors propose that 2AO can be formed from cyanamide (NH$_2$CN) and glycolaldehyde (HOCH$_2$CHO), under plausible early Earth conditions. Since both precursors have been detected in the interstellar medium (Turner et al. 1975; Hollis et al. 2000), the presence of 2AO in astrophysical environments cannot be discarded. It is generally assumed that complex organic molecules in space are most likely formed either at the surface or in the bulk of ice layers. The present work provides IR signatures of solid 2AO at low temperature, that can help to search for this species in ices, both in cold Solar System objects or in dense clouds in the interstellar medium. In particular, we believe that the spectral region presented in Figure 4, between 1900 cm$^{-1}$ and 660 cm$^{-1}$ (5.36-16.67 μm), where characteristic narrow absorptions are present, would be most adequate for astronomical searches. However, the probability of detecting this molecule in the ices is low, considering that it would be present in a small proportion, and that other species might interfere. Detection in the gas phase, where highly selective radioastronomic techniques are used, would be easier but, as mentioned above, the first attempt has been unsuccessful (Jiménez-Serra et al. 2020), which raises the question of whether this molecule is not formed in significant amounts or whether it is destroyed before reaching the gas phase.

In the previous sections we have studied the destruction of 2AO ices with UV photons or 5 keV electrons, which were intended to mimic the effect of UV fields and cosmic rays on astronomical ice analogues. We can now take these experimental results to estimate the stability (half-life) of the molecule in the ice mantles of dust grains in dense interstellar clouds, or in icy bodies in the outer Solar System. To do so, we use literature values for the energy received by the ices in these environments, and half-life doses for 2AO ices derived from the experimental measurements in this work. The results are listed in Table 9. Given the assumptions and approximations made, and the uncertainty in the experimental data, we can only obtain an order-of-magnitude estimate.

Complex organic molecules, if present at all, will be minor components of astronomical ices. Observations show that the ices are mostly made $H_2O$ with smaller amounts of CO, $CO_2$, $CH_3OH$, $CH_4$, and $NH_3$. For the present stability estimates we have just considered binary mixtures of 2AO with $H_2O$, with water being the major component. For the estimation of the effect of the UV light we have taken a mixture of 2AO (5%) in water. This mixture was not directly studied in the experiments, but its half-life dose can be estimated with equation (7). As commented above, the UV destruction cross section of 2AO is not much affected by the dilution in water ice. The main effect of the water molecules in the ice mixture is just the absorption of a part of the incoming radiation. Using the measured destruction cross section for 2AO ($\sigma_{des} \approx 9.5 \times 10^{-18}$ cm$^2$) and the absorption cross section for the ice mixture ($\sigma_a = 4.1 \times 10^{-18}$ cm$^2$), one gets a half-life dose of 51 eV/molec.

Table 9. Half-life of 2-aminooxazole ice in astrophysical environments. The UV half-lives were calculated with a 2AO UV half-life dose of 51 eV/molec, which corresponds to the value given by equation (7) for a 2AO (5%)/$H_2O$ ice mixture (see text). The Cosmic Ray half-lives were calculated with a half-life dose of 43 eV/molec (E2 sample in Table 7), which corresponds to the a 2AO (22%)/$H_2O$ ice mixture.

| Astrophysical environment | Life time of ices (yr) | UV Dose Rate (eV molec$^{-1}$ yr$^{-1}$) | 2-aminooxazole UV Half-life (yr) |
|---|---|---|---|
| Kuiper Belt Object | 4.6 x 10$^9$ | 2.2 x 10$^{-2}$ (a) | 2.3 × 10$^3$ (e) |
| Cold Dense Cloud | 10$^7$ | 4 x 10$^{-7}$ (b) | 1.3 x 10$^8$ |
|  |  | CR Dose Rate (eV molec$^{-1}$ yr$^{-1}$) | 2-aminooxazole CR Half-life (yr) |
| Kuiper Belt Object | 4.6 x 10$^9$ | 5.6 x 10$^{-3}$ (c) | 7.7 x 10$^3$ |
| 40 Au, 30 μm depth |  | 1.6 x 10$^{-8}$ (d) | 2.7 x 10$^9$ |
| Cold Dense Cloud | 10$^7$ | 3 x 10$^{-7}$ (b) | 1.4 x 10$^8$ |

(a) Moore & Hudson (2005), UV dose rates are estimated for the top 15 nm of the ice. (b) Moore et al. (2001), UV dose are estimated for typical ice mantles with thickness of 20 nm. (c) Cooper et al.

(2003), CR dose rate for ice thickness lower that 10 nm. (d) Strazzulla et al. (2003). (e) This value of the UV half-life is actually an upper limit, since the UV half-life doses of this work and the UV dose rate used for the calculation correspond to VUV photons and 2AO can undergo photolysis with longer wavelength photons, which are abundant in the Solar System (see text).

For the assessment of the stability of 2AO toward cosmic rays we have used the results of the 5keV electron bombardment on the 2AO (22%)/water mixture (sample E2 in Tables 4, and 7). The corresponding half-life dose is 43 eV. Note that in this case the half-life dose hardly changes with water dilution since the release of energy to the ice sample is more indiscriminate. The propagation of the secondary electron cascade during electron bombardment depends mostly on the type and density of atoms in the solid and is not so sensitive to the specific molecular species.

Table 9 shows that in the ice mantles of cold, dense clouds, 2AO can persist under UV and CR irradiation for more than $10^8$ years. This is a promising finding, since it implies that, if the molecule were present in the ice mantles of dust grains, it could survive the cloud collapse and participate in further chemical networks in protoplanetary disks. Moreover, if formed in a large enough amount, it could be observed in chemically rich astronomical sources such as massive hot cores, low-mass warm-cores (or hot corinos) or Galactic Center Giant Molecular Clouds (GMCs) or maybe even in cold cores (Jimenez-Serra et al. 2016). However, a previous search of this prebiotic species toward a hot corino and a Galactic Center GMC in the gas phase in the millimetre wavelength range, did not yield any detection (Jimenez-Serra et al. 2020). The determination of the absorption spectra of 2-aminooxazole in this work will enable searches in the solid phase with the James Webb Space Telescope.

On the ice surface of Kuiper Belt Objects (KBOs), UV photons and CRs will destroy 2AO very quickly, in just thousands of years. Any 2AO hypothetically formed in the solar nebula would disappear from the surface of KBOs at the beginning of the evolution of the Solar System. In fact, the UV half-life time given in Table 9 is an upper limit, since both the destruction cross section measured in this work, and the dose rate used for the evaluation correspond to photons with wavelengths < 180 nm, which are only adequate for the interstellar medium, where the UV field is limited to these wavelengths, or for the most common small molecules in the Solar System, that do not absorb at longer wavelengths. However, UV absorption in 2AO is intense down to 240 nm and for unscreened solar radiation (Tobiska et al. 2000) the proportion of photons with λ in the 180-240 nm range is large. The low photostability of 2AO under mid-range UV light (210-290 nm) was stressed by Todd et al. (2019). These authors, who studied the photodegradation of this molecule in an aqueous solution, concluded that its half-life would be just a few hours for the UV flux expected at the surface of the early Earth. The present photolysis data, which extend those of Todd et al. (2019) toward the VUV (180-120 nm) range, show that the molecule would be also short lived, in evolutionary terms, at the surface of outer Solar System bodies.

The survival of fragile molecules can be much increased if they are protected by layers of ice or refractory materials. Usually, a thin ice layer is enough to protect molecules from the external UV field or from CRs, which in the Solar System contain a large amount of slow protons from the solar wind with a small penetration depth (Cooper et al. 2003). Table 9 shows that an ice layer of just 30 μm, can increase the survival time of 2AO under CR bombardment by more than four orders of magnitude. However, water ice is not a good shield against the UV photolysis of 2AO. Although the UV field for photons with λ < 160 nm disappears virtually a few microns below the surface, the absorption cross section of water ice drops abruptly beyond 160 nm and the solid becomes essentially transparent for λ > 180-200 nm (Warren and Brandt 2008) where 2AO still undergoes strong absorption (see Fig. 8) and photolysis. Still, the molecule could be effectively protected under rocky material, but plausible scenarios for prebiotic chemistry require the presence of water. The survival probability of 2 AO would be enhanced if the ice covered grains, where the molecule can form, agglomerate quickly enough to from clusters and cometesimals, where the ice will be protected from UV radiation.

5. Summary and conclusions

The IR spectrum of amorphous and crystalline 2-aminooxazole ice was experimentally studied in the 4000-600 $cm^{-1}$ range. Theoretical calculations performed on a tentative solid structure produced spectra that were in fair agreement with the measurements and allowed the assignment of the absorption bands. The most intense bands are associated with $NH_2$ stretching (≈ 3500-3000 $cm^{-1}$) and bending (≈ 1900-1500 $cm^{-1}$) vibrations. Ring vibrations and bands of various deformation modes appear between ≈ 1450 and 700 $cm^{-1}$. Band strengths were derived from polycrystalline 2AO samples in KBr pellets and were used for the estimate of column densities and ice layer thicknesses. The region between 1900 and 700 $cm^{-1}$ presents characteristic absorptions for the molecule that could be eventually used for searches in astronomical ices with the James Webb Space Telescope..

The UV absorption spectrum of solid 2AO was measured in the 120-250 nm interval. Over this range, the molecule presents continuum absorption with a maximum at 158 nm and broader maxima at larger wavelengths.

Ice samples of 2-aminooxazole and of mixtures of 2AO with water were subjected to irradiation by UV (120-180 nm) photons and to bombardment by 5 keV electrons to simulate the effects of UV fields and cosmic rays in space. IR spectra, recorded for all samples after energetic processing, were used to study the reaction products. In general, individual molecules could not be specified just with the IR spectra, but they allowed the identification of functional groups. The main photoproducts were found to be similar in the UV and electron processing experiments. $CO_2$ (2423 $cm^{-1}$), CO (2137 $cm^{-1}$) and $OCN^-$ (2164 $cm^{-1}$) were found in all cases. NH stretching bands of amines and/or amides appeared beyond 3000 $cm^{-1}$. Features associated with cyanates, isocyanates, nitriles and possibly HCN were seen in the 2280-2000 $cm^{-1}$ interval. A broad absorption band with a maximum at ≈ 1700 $cm^{-1}$ (C=O) and various secondary maxima or shoulders toward lower wavenumbers, extending down to ≈ 1200 $cm^{-1}$, were also observed. Some of the maxima in this broad feature were consistent with amide bands (I to III) and suggest the formation of some kind of polymer. Upon heating to 200 K, the profile of the 1700-1200 $cm^{-1}$ band was somewhat modified but a polymeric residue remained.

Destruction cross sections for 2-aminooxazole ice, pure and in mixtures with water, under irradiation with UV (6.3-10.9 eV) photons or 5 keV electrons were obtained from the decay of the 1423 cm$^{-1}$ band of 2AO, which is associated with an OCN stretching vibration of the ring. The destruction cross section for UV irradiation was found to be comparatively high ($\sigma_{des} \approx 9.5 \times 10^{-18}$ cm$^2$) and did not change appreciably in the mixtures with water. Water molecules absorb in the UV range studied and their only effect seems to be the reduction of the photon flux available for the destruction of 2AO. In contrast, in the electron bombardment experiments, the destruction cross section increases upon dilution of 2AO in water, which suggests that the reactive species generated in the cascade of secondary electrons within the ices contribute to the destruction of the molecule.

The measured destruction cross sections were used to calculate half-life energy doses for the various samples. Using these half-life doses we have estimated the expected half-life times of 2-aminoxazole in the ices of dense clouds and at the surface of Kuiper belt objects. To do so, we took literature values for the UV and CR fluxes in these environments. In the ice mantles of dense clouds, 2AO should be stable against UV radiation and cosmic rays, with half-life times of $\approx 10^8$ yr, longer than the typical life of these clouds ($\approx 10^7$ yr). It could thus survive the cloud collapse and, if formed in a sufficient amount through ice chemistry, it could be observable protostars and maybe in prestellar cores. In contrast, 2AO would be very unstable at the surface of Kuiper belt objects, where it would be destroyed in just thousands of years either by CRs or by UV photons, like those used in this work, unless protected by a layer of ice or of refractory materials. Furthermore, the stability of 2-aminooxazole in ices within the Solar System would be much reduced by the large absorption and photodissociation of the molecule beyond 200 nm. For wavelengths longer than 180-200 nm, water ice becomes transparent and offers no protection against the intense UV solar radiation. In spite of its low photostability, the molecule could still act as an intermediate in prebiotic synthesis in Solar System objects, but only in environments protected by rocky or carbonaceous materials, like the interior of comets.

Acknowledgements

BM, VT, IT, and VJH are grateful to the Ministerio de Economia y Competitividad (MINECO) of Spain under grant FIS2016-77726-C3-1-P. GMMC, HC, and CGD acknowledge MINECO support under grant AYA2017–85322-R (AEI/FEDER, UE), Ph.D. fellowship FPU-17/03172, and MDM-2017–0737 Unidad de Excelencia 'María de Maeztu' – CAB (CSIC-INTA). IJS acknowledges partial support from the Spanish FEDER (ESP2017-86582-C4-1-R) and the State Research Agency (PID2019-105552RB-C41).